\documentclass{article}

\usepackage{arxiv}

\usepackage[utf8]{inputenc} 
\usepackage[T1]{fontenc}    
\usepackage{hyperref}       
\usepackage{url}            
\usepackage{booktabs}       
\usepackage{amsfonts}       
\usepackage{nicefrac}       
\usepackage{microtype}      
\usepackage{cleveref}       
\usepackage{lipsum}         
\usepackage{graphicx}
\usepackage{natbib}
\usepackage{doi}

\title{Un sistema automatizado para el adiestramiento y evaluaci\'on
  de estudiantes de f\'isica ante ex\'amenes cerrados.}

\date{Febrero, 2023}

\author{ \href{https://orcid.org/0000-0001-6886-2182}{\includegraphics[scale=0.06]{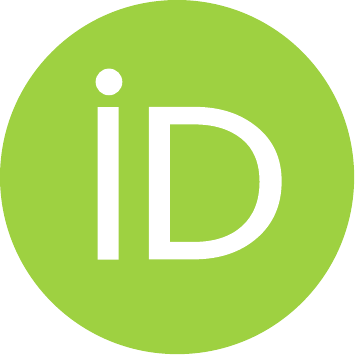}\hspace{1mm}Juan Jos\'e Jim\'enez-Torres}\\
	Kanazawa Institute of Technology\\
	Yatsukaho Research Center\\
	Hakusan, Ishikawa 924-0838, Japan \\
	\texttt{JuanJzTorres@gmail.com} \\
}

\renewcommand{\shorttitle}{Research Working Paper}

\begin{document}
\maketitle

\begin{abstract}

In this work, I present an automatic system for the evaluation of
closed-type exercises in physics at the high school level or in the
first year of a degree where physics is a mandatory course. It is
expected that this will allow students or teachers with access to a
computer or mobile device to get into a repository of exercises for
different areas of physics and practice them using true or false or
multiple choice options.

This system is expected to solve the problem that many students do not
have enough skills to face this type of evaluation, since they have
not had the opportunity to work with closed-type exams that are graded
by computers and are used by educational institutions such as
evaluation tools. This material can also benefit teachers by providing
them with an alternative form of evaluation to the traditional form.

To use this system, computer programming skills are not required,
since it consists of a program where the answers are selected and the
program provides a numerical score, a brief interpretation, and the
solutions to every exercise, which will allow students to compare
their answers with the solutions.

\bigskip

\bigskip

{\centerline{\bf RESUMEN}}

En este trabajo presento un sistema automatizado para la evaluaci\'on
de ejercicios cerrados de f\'isica de nivel bachillerato o de primer
a\~no de licenciatura en donde la f\'isica sea un curso
obligatorio. Se espera que esto permita a alumnos o docentes con
acceso a una computadora o un dispositivo m\'ovil acceder a un
repositorio de ejercicios de distintas \'areas de la f\'isica y
practicarlos mediante las opciones de falso o verdadero o de opci\'on
m\'ultiple.

Este sistema pretende ayudar a la soluci\'on del problema referente a
que muchos alumnos no tienen las habilidades para enfrentarse a ese
tipo de evaluaciones, pues no han tenido las oportunidades de trabajar
con ex\'amenes cerrados que son calificados por computadoras y que son
utilizados por instituciones educativas como herramientas de
evaluaci\'on. Este material puede beneficiar tambi\'en a docentes al
brindarles una forma de evaluaci\'on alternativa a la forma
tradicional.

Para usar este sistema no se requieren habilidades de programaci\'on
en c\'omputo, pues consiste de un programa en donde se seleccionan las
respuestas y el programa entrega una calificaci\'on num\'erica, una
breve interpretaci\'on y las soluciones a cada uno de los ejercicios,
lo cual permitir\'a a los estudiantes confrontar sus respuestas con
las soluciones.

\end{abstract}



\section{Introducci\'on}\label{introduccion}

Hoy en d\'ia la evaluaci\'on educativa corresponde a un t\'opico
fundamental en la investigaci\'on educativa, considerando que aunque
se cuenten con nuevas herramientas tecnol\'ogicas para la ense\~nanza,
si no se adapta la forma de evaluaci\'on a los tiempos actuales, el
proceso de ense\~nanza de las ciencias se encontrar\'a
incompleto. Esto significa que las evaluaciones no deber\'ian
concentrarse \'unicamente en la calificaci\'on de ex\'amenes a l\'apiz
en la forma tradicional. Adem\'as de esto, nos encontramos con que
actualmente los ex\'amenes cerrados son muy frecuentes y que no todos
los alumnos tienen las habilidades para enfrentarse a este tipo de
evaluaciones, pues no todos los alumnos han tenido una oportunidad de
enfrentarse a este tipo de ex\'amenes automatizados.

En este trabajo, se presenta un programa interactivo de evaluaci\'on
de ejercicios de f\'isica que est\'a construido en base a preguntas de
opci\'on m\'ultiple, y de falso o verdadero, las cuales son preguntas
del tipo que pueden encontrarse en los ex\'amenes cerrados y
calificados autom\'aticamente por una computadora. As\'i mismo, forma
parte de una serie de trabajos para el desarrollo de software para la
ense\~nanza y evaluaci\'on de la f\'isica, iniciados con el art\'iculo
de Jim\'enez-Torres et al. (2018).

El programa interactivo contiene un banco de reactivos para las
asignaturas de f\'isica, las cuales abarcan temas de f\'isica
cl\'asica, una breve introducci\'on a la F\'isica Moderna y una
asignatura para el tema de la f\'isica y su relaci\'on con otras
ciencias. Espec\'ificamente estas asignaturas son: 1) La F\'isica y su
relaci\'on con otras ciencias naturales, 2) Fen\'omenos Mec\'anicos,
3) Fen\'omenos Termodin\'amicos, 4) Fen\'omenos Ondulatorios, 5)
Fen\'omenos electromagn\'eticos, y 6) F\'isica contempor\'anea:
Relatividad Especial e Introducci\'on a la Mec\'anica Cu\'antica.

Este material est\'a planeado para cursos introductorios o
diagn\'osticos y se puede emplear para un curso de una duraci\'on de
un semestre universitario. Debido a la naturaleza del contenido, se
recomienda que el estudiante cuente con conocimientos b\'asicos de
matem\'aticas y f\'isica. Para cada una de las asignaturas presentadas
en este trabajo, se presentan ejercicios y problemas resueltos que
proporcionan herramientas y t\'ecnicas necesarias para afrontar los
problemas que los estudiantes pueden encontrar en sus tareas y
ex\'amenes. Para estos ejercicios, he empleado el sistema
internacional de unidades para la soluci\'on de ejercicios.

Por otra parte, para apoyar a lograr los objetivos ideales de egreso
de las asignaturas o cursos de f\'isica del bachillerato, se
considerar\'an los distintos niveles cognoscitivos indicados en el
N\'ucleo de conocimientos y formaci\'on b\'asicos que debe
proporcionar el Bachillerato de la UNAM. Este consiste en un documento
interno y oficial del Colegio de Ciencias y Humanidades (CCH) y de la
Escuela Nacional Preparatoria (ENP), ambos parte de la Universidad
Nacional Aut\'onoma de M\'exico (UNAM). Estos niveles cognoscitivos se
encuentran clasificados de acuerdo a: posesi\'on de informaci\'on,
comprensi\'on, elaboraci\'on conceptual y soluci\'on de problemas. De
acuerdo a los contenidos de las unidades en los cursos de f\'isica y
los objetivos a alcanzar, en este trabajo se establece una
ponderaci\'on al n\'umero de preguntas que debe elaborarse en cada
nivel cognoscitivo.

As\'i mismo, se espera que este programa interactivo no solo sea una
valiosa fuente de informaci\'on y adiestramiento para estudiantes de
bachillerato o licenciatura, sino que tambi\'en pueda servir de apoyo
y actualizaci\'on para los docentes de los cursos de f\'isica. Todo
esto es desarrollado con la finalidad de contribuir en el mejoramiento
del nivel acad\'emico en la educaci\'on media superior y superior.

\section{Justificaci\'on}\label{justificacion}

En los cursos de f\'isica de nivel de bachillerato se estudian
principalmente los conceptos, de los cuales algunos son f\'aciles de
reconocer debido a la experiencia diaria de los alumnos, y otros por
su parte son m\'as complejos y abstractos y los estudiantes
\'unicamente los pueden identifican despu\'es de que han sido
derivados detalladamente por el profesor (Clarence 1979). La f\'isica
es una ciencia que no solo trata con la causa o raz\'on de los
fen\'omenos naturales, sino que tambi\'en se enfoca en determinar
cantidades y magnitudes, y por esta raz\'on se involucran problemas de
naturaleza num\'erica en los cursos de f\'isica. Algunos autores como
Seville (1959) y Caballer y O\~norbe (1999) han considerado que un
m\'etodo efectivo en la ense\~nanza de la f\'isica consiste en
trabajar con problemas o ejercicios de tipo num\'erico y recomienda
realizarlo de manera sistem\'atica y adecuada. Por esta raz\'on, los
estudiantes de f\'isica deben encontrarse capacitados para resolver
problemas o ejercicios formulados en lenguaje matem\'atico.

Sin embargo, muchos alumnos encuentran la soluci\'on de problemas como
una forma de evaluaci\'on dif\'icil, pues muchos de ellos consideran
que la soluci\'on de ejercicios consiste \'unicamente en sustituir
valores en una f\'ormula. Esta suposici\'on muchas veces tiene un
resultado desfavorable para sus calificaciones obtenidas. El programa
interactivo se presenta como una herramienta auxiliar para este
problema, y que permita al alumno adquirir habilidades para la
soluci\'on de ejercicios de f\'isica y aplicaci\'on de sus
conocimientos adquiridos en sus cursos.

Una vez mencionada la importancia de los problemas y ejercicios en los
cursos y ex\'amenes de f\'isica, debemos considerar que hay varias
razones m\'as para considerar a una computadora o dispositivo
electr\'onico como instrumentos para la educaci\'on, particularmente
en la ense\~nanza de la f\'isica en los niveles de bachillerato y
universitario (Calder\'on 1988). Hoy en d\'ia, los sistemas
automatizados de evaluaci\'on en las ciencias realizan un papel
valioso en los procesos de evaluaci\'on en los sistemas educativos (UN
2011). Los sistemas interactivos automatizados se han empleado desde
hace d\'ecadas para la evaluaci\'on masiva de ex\'amenes de
estudiantes que solicitan ingreso a las universidades p\'ublicas y
privadas en todo el mundo. De acuerdo a la agenda de las Naciones
Unidas para la Educaci\'on, la inclusi\'on de la tecnolog\'ia en los
sistemas de educaci\'on tiene claras evidencias de impacto positivo,
particularmente en los pa\'ises en v\'ias de desarrollo, en donde
varios pa\'ises reportan una infraestructura inadecuada para la
aplicaci\'on de ex\'amenes a una gran cantidad de estudiantes de
manera simult\'anea.

Por otra parte, la carencia de profesores capacitados en las \'areas
cient\'ificas representa un obst\'aculo en el desarrollo del proceso
educativo. Incluso en lugares donde se cuenta con una planta docente
suficiente, existen algunos de estos profesores que declaran no contar
con habilidades o competencias para la implementaci\'on de ex\'amenes
cerrados automatizados para sus estudiantes. En este sentido,
Leiserson y Vinney (2015) recomiendan que los profesores deben
tambi\'en estar actualizados en el manejo y dominio de las nuevas
tecnolog\'ias durante sus carreras acad\'emicas y no solamente los
estudiantes.

En la literatura y distintos sitios en Internet existen ejercicios de
ex\'amenes cerrados de f\'isica de opci\'on m\'ultiple o de falso o
verdadero, aunque la mayor\'ia de ellos se encuentran en idioma
ingl\'es, lo que representa una barrera para varios
estudiantes. Tambi\'en existen sitios de Internet donde ofrecen
ejemplos de ejercicios de f\'isica con respuestas de opci\'on
m\'ultiple o de falso o verdadero, sin embargo la cantidad de estos
ejercicios puede ser insuficiente para cubrir todos los temas de las
distintas \'areas de la f\'isica. Por su parte en otros sitios se
ofrecen ejercicios de f\'isica de opci\'on m\'ultiple o de falso o
verdadero, en donde \'unicamente se limitan a entregar una
calificaci\'on, sin mostrar la f\'isica de la soluci\'on correcta de
los ejercicios, lo que puede dejar al estudiante sin saber porqu\'e su
respuesta seleccionada es incorrecta cuando obtiene una calificaci\'on
desfavorable. Por lo que la implementaci\'on de sistemas automatizados
y actualizados de evaluaci\'on puede resultar en un gran apoyo para la
soluci\'on de estas problem\'aticas.

Otra problem\'atica radica en que algunos estudiantes que logran
resolver un problema de f\'isica de manera correcta, lo hacen gracias
al empleo de conocimientos memorizados, pero que cuentan con poca
capacidad de entender o resolver problemas de alg\'un otra \'area
distinta de la f\'isica. Esta situaci\'on puede tener consecuencias
notorias en un nivel de educaci\'on superior, cuando los estudiantes
se enfrentan a un problema de f\'isica y argumentan que comprenden el
concepto de la f\'isica, pero que no encuentran la forma para poder
resolver los problemas de un examen cerrado con l\'imite de
tiempo. Esta problem\'atica es una motivaci\'on m\'as para el
desarrollo e implementaci\'on de nuevos m\'etodos educativos para
atender esta situaci\'on. Por lo que el sistema presentado en este
trabajo puede ayudar a fortalecer la capacidad y habilidad para la
soluci\'on de ejercicios cerrados.

El sistema automatizado de evaluaci\'on en este trabajo tambi\'en
puede ayudar a aquellos estudiantes que no pueden atender a clases de
manera presencial por carencia de recursos o por situaciones
espec\'ificas como la experimentada durante el brote de pandemia de
SAR COVID 2019 que oblig\'o a un confinamiento en el hogar para los
estudiantes y la implementaci\'on de educaci\'on en l\'inea para los
estudiantes y profesores.

\section{Taxonom\'ia}\label{taxonomia}

En esta secci\'on se mencionan los niveles cognoscitivos indicados por
el N\'ucleo de Conocimientos y Formaci\'on B\'asicos (NCFB) que debe
proporcionar el Bachillerato de la UNAM (ECUNAM 2001) y en los que se
encuentran basados los ejercicios y problemas de f\'isica de este
trabajo. El NCFB consiste en un documento elaborado por el consejo
acad\'emico del bachillerato de la UNAM, en donde se plantean los
conocimientos, habilidades, y actitudes prioritarios y de mayor
relevancia para obtener los conocimientos ideales de egreso de los
estudiantes de bachillerato de la UNAM en M\'exico. El NCFB pretende
que la educaci\'on adquirida por los estudiantes no est\'e sometida
\'unicamente a la adquisici\'on de informaci\'on, sino que fomente en
ellos el pensamiento anal\'itico y cr\'itico, y el desarrollo de
habilidades para la soluci\'on de problemas.

Para la elaboraci\'on de los objetivos cognoscitivos, el NCFB emplea
los siguientes 4 niveles cognoscitivos:

I. Nivel de posesi\'on de informaci\'on. En este nivel, el alumno
\'unicamente recuerda y reproduce la informaci\'on adquirida en sus
cursos.

II. Nivel de comprensi\'on. En este nivel, el alumno entiende e
interpreta la informaci\'on adquirida en sus cursos, sin alterar el
significado o definici\'on original de esa informaci\'on.

III. Nivel de elaboraci\'on conceptual. En este nivel, el alumno
desarrolla procesos de an\'alisis, s\'intesis y evaluaci\'on de la
informaci\'on adquirida en sus cursos.

IV. Nivel de soluci\'on de problemas. En este nivel, el alumno emplea
el conocimiento adquirido y sus habilidades de comprensi\'on y
razonamiento de un tema para el desarrollo de una secuencia de pasos
para lograr una soluci\'on de un ejercicio o problema.

Estos niveles cognoscitivos tambi\'en son descritos en textos como
Fowler (2002) y Woolfolk (1996).

Los reactivos presentados en este trabajo contienen ejercicios que
incluyen estos niveles cognoscitivos. De esta manera se pretende que
los alumnos comprendan la f\'isica involucrada en los fen\'omenos
f\'isicos que ocurren en la naturaleza y que los problemas
relacionados a ellos sean no solo de car\'acter emp\'irico donde
solamente se aplique de forma mec\'anica expresiones matem\'aticas,
sino que tambi\'en sean de car\'acter anal\'itico y cuantitativo.

Los objetivos educativos del \'area de f\'isica se desarrollan
pensando en los requerimientos y necesidades de las carreras a nivel
licenciatura que requieren conocimientos de f\'isica como las
ingenier\'ias. Adem\'as, estos objetivos est\'an pensados para que los
alumnos adquieran conocimientos y habilidades que le sean de utilidad
para aplicarlos en la vida diaria, y que de esta manera tengan un
inter\'es aut\'entico por el estudio de la f\'isica. En el NCFB de la
UNAM, promueven el desarrollo de habilidades superiores a aquellas
relacionadas a la memorizaci\'on o a la aplicaci\'on mec\'anica de
algoritmos, o f\'ormulas matem\'aticas.

\section{Metodolog\'ia a seguir para la elecci\'on de los ejercicios}\label{metodologia}

En el proceso de evaluaci\'on de la f\'isica es importante la
obtenci\'on de informaci\'on del conocimiento que poseen los
estudiantes, y esto se puede lograr mediante la elaboraci\'on y
aplicaci\'on de medios de evaluaci\'on como lo son las t\'ecnicas, los
instrumentos y los ejercicios. En las t\'ecnicas se dirige una
actividad y se busca el aprovechamiento de los recursos o materiales
existentes en t\'erminos pr\'acticos. Un instrumento consiste en un
medio para lograr un objetivo y constituyen a las t\'ecnicas. Por su
parte, los ejercicios son los componentes del instrumento y analizan
una situaci\'on f\'isica o problema que requiere una soluci\'on (CB
1994).

Las t\'ecnicas de evaluaci\'on que se emplean frecuentemente se
refieren a: I) soluci\'on de problemas; II) solicitud de productos;
III) observaci\'on; y IV) interrogatorio.

En la soluci\'on de problemas se eval\'uan los conocimientos del
estudiante a nivel conceptual o de compresi\'on, tambi\'en se analiza
el conjunto de desarrollos para llegar a una soluci\'on.

En la solicitud de productos se requiere que el estudiante presente
documentos escritos, dise\~nos o montajes experimentales relacionados
a un conocimiento adquirido en las aulas.

En la observaci\'on se observa y documenta las expresiones referentes
al conocimiento del estudiante sobre un tema espec\'ifico.

Y en el interrogatorio se solicita informaci\'on donde se registran
las respuestas directas, interpretaciones de la soluci\'on a un
problema o la opini\'on del estudiante ante un problema.

En este trabajo \'unicamente se incluyen ejercicios para la t\'ecnica
de soluci\'on de problemas I) y los instrumentos incluidos en esta
t\'ecnica son las i) pruebas objetivas, ii) los desarrollos
tem\'aticos y iii) los simuladores escritos.

Los desarrollos tem\'aticos corresponden a un instrumento de
producci\'on o generaci\'on, mientras que las pruebas objetivas y
simuladores escritos corresponden a instrumentos de selecci\'on
(Hern\'andez 2002; Wood 2000), los cuales son la base fundamental para
el desarrollo de los reactivos de f\'isica en este trabajo.

En los ejercicios de los instrumentos de selecci\'on los reactivos
tienen respuestas \'unicas. Pueden ser aquellos con una respuesta
corta, complementaci\'on de un enunciado, respuestas de falso o
verdadero o ejercicios de opci\'on m\'ultiple. En este trabajo
\'unicamente presentar\'e ejercicios de f\'isica de opci\'on
m\'ultiple y con respuestas de falso o verdadero.

Los ejercicios de opci\'on m\'ultiple consisten de un enunciado
principal en donde se expresa una situaci\'on o un enunciado referente
a un fen\'omeno o principio de f\'isica e incluyen un conjunto de
alternativas de respuestas, donde s\'olo una es la respuesta correcta
y se encuentran distribuidas al azar para todos los ejercicios en este
trabajo. Este tipo de reactivos permitir\'a evaluar distintos niveles
del conocimiento del estudiante, desde la posesi\'on de informaci\'on
hasta el an\'alisis de situaciones complejas en un problema de
f\'isica. Para motivar el inter\'es de los estudiantes en esta
evaluaci\'on, he elaborado ejercicios de f\'isica con aplicaciones en
la vida diaria que les resulten atractivos y no abstractos o de
dif\'icil comprensi\'on. En todos los ejercicios de opci\'on
m\'ultiple y de falso o verdadero he planteado problemas que cubran
los requerimientos de los objetivos de aprendizaje planteados en la
secci\'on de taxonom\'ia.

En los reactivos o ejercicios de falso o verdadero se muestra una
proposici\'on o contexto f\'isico donde el estudiante debe comprender
si ese contexto planteado es correcto o falso y seleccionar una
opci\'on como correcta. El orden de las opciones de respuestas
correctas han sido colocadas al azar para todos los ejercicios.

El repositorio de ejercicios en este trabajo est\'a elaborado de
manera que la cantidad de reactivos de falso o verdadero sea inferior
a los ejercicios de opci\'on m\'ultiple, ya que en los ejercicios de
falso o verdadero existe la posibilidad de que el estudiante
respondiendo de manera azarosa, pueda tener una posibilidad del 50
$\%$ de seleccionar la respuesta correcta. Los reactivos y ejercicios,
y sus respectivas figuras, se encuentran basados en ejercicios
planteados en distintas fuentes bibliogr\'aficas como los textos de
Schad (1996), Resnick y Halliday (1981), Tippens (2001), Benson (1999,
1991) y Hewitt (1999). Estos ejercicios de f\'isica tambi\'en fueron
empleados para el trabajo que desarrolle en Jim\'enez-Torres (2006).

\section{Sistema Automatizado}\label{sistema}

En este trabajo presento un sistema automatizado interactivo de
evaluaci\'on de ejercicios de f\'isica donde una computadora es
empleada por un estudiante. En este sistema la interfaz y el c\'odigo
fuente est\'an escritos en el lenguaje de programaci\'on HTML. Este
sistema est\'a dise\~nado para que el estudiante \'unicamente se
enfoque en responder las preguntas y no pierda tiempo en ning\'un paso
de programaci\'on en c\'omputo. Los estudiantes seleccionan con un
clic la opci\'on de la respuesta que ellos consideran es la correcta
frente a cada ejercicio que encuentran en el repositorio de ejercicios
de f\'isica. Despu\'es, cuando han seleccionado todas sus respuestas,
dan clic en el bot\'on calificar y el sistema les muestra
autom\'aticamente una calificaci\'on num\'erica en el rango desde 0 a
10, una breve interpretaci\'on de su respuesta obtenida, y una
propuesta para la soluci\'on de cada uno de los ejercicios que
encontr\'o en el sistema.

La breve interpretaci\'on consiste en un mensaje corto del tipo:
INSUFICIENTE, SUFICIENTE, REGULAR, BIEN, MUY BIEN, o EXCELENTE,
dependiendo de su calificaci\'on obtenida. Estas interpretaciones
est\'an basadas en el sistema de calificaciones de la educaci\'on en
M\'exico (SEP 1978), donde una calificaci\'on de 0 a 5.99 corresponde
a una interpretaci\'on de INSUFICIENTE; 6 a 6.49 a una
interpretaci\'on de SUFICIENTE; 6.5 a 7.49 a una interpretaci\'on de
REGULAR; 7.5 a 8.49 a una interpretaci\'on de BIEN; 8.5 a 9.49 a una
interpretaci\'on de MUY BIEN; y 9.5 a 10 a una interpretaci\'on de
EXCELENTE.

Este sistema cuenta actualmente con 245 ejercicios en total, donde 60
de ellos corresponden al \'area de mec\'anica, 42 al \'area de
termodin\'amica, 53 al \'area de electromagnetismo, 39 al \'area de
relatividad y mec\'anica cu\'antica, 33 al \'area de fen\'omenos
ondulatorios y 18 a la relaci\'on de la f\'isica con otras ciencias
naturales. En las siguientes direcciones se muestran las estructuras
del sistema, las preguntas de los reactivos, y las opciones
m\'ultiples de las respuestas para cada una de las \'areas de la
f\'isica que contiene el sistema. Despu\'es de que el usuario ha dado
clic en el bot\'on calificar, aparecen propuestas para las soluciones
de cada uno de los ejercicios que contiene el sistema.

Ejercicios de {\it La relaci\'on de la f\'isica con otras ciencias
  naturales}:
\href{https://juanjztorres.github.io/ejca/}{https://juanjztorres.github.io/ejca/}

Ejercicios de {\it Mec\'anica}:
\href{https://juanjztorres.github.io/ejme/}{https://juanjztorres.github.io/ejme/}

Ejercicios de {\it Termodin\'amica}:
\href{https://juanjztorres.github.io/ejte/}{https://juanjztorres.github.io/ejte/}

Ejercicios de {\it Fen\'omenos ondulatorios}:
\href{https://juanjztorres.github.io/ejon/}{https://juanjztorres.github.io/ejon/}

Ejercicios de {\it Electromagnetismo}:
\href{https://juanjztorres.github.io/ejel/}{https://juanjztorres.github.io/ejel/}

Ejercicios de {\it Relatividad y mec\'anica cu\'antica}:
\href{https://juanjztorres.github.io/ejcu/}{https://juanjztorres.github.io/ejcu/}

\section{Conclusi\'on}\label{conclusion}

En este sistema, present\'o un repositorio de ejercicios resueltos de
f\'isica donde los estudiantes o profesores pueden disponer de un
instrumento de evaluaci\'on de conocimientos de f\'isica para nivel
bachillerato o primer a\~no de nivel universitario a nivel
licenciatura. Este sistema permite conocer no s\'olo una
calificaci\'on num\'erica, sino tambi\'en una propuesta para conocer
la soluci\'on correcta a cada uno de los ejercicios de este
repositorio. Present\'o ejercicios para la evaluaci\'on de las
distintas \'areas de la f\'isica como la mec\'anica, termodin\'amica,
\'optica, electromagnetismo, relatividad especial y mec\'anica
cu\'antica, las cuales est\'an basados en las unidades de f\'isica que
proporciona el bachillerato universitario de la UNAM. Todos los
ejercicios cuentan con soluciones entregadas por el sistema para que
el alumno identifique las discrepancias en sus respuestas.

Se espera que su impacto sea favorable, adem\'as de que se
encontrar\'a abierto para futuras ediciones o mejoramientos donde un
docente o experto puede adicionar m\'as material educativo para que de
esta manera se disponga de un mayor repositorio de ejercicios
resueltos de f\'isica y que sea de utilidad no solo para estudiantes
de las licenciaturas de f\'isica o de ingenier\'ias, sino tambi\'en de
estudiantes de otras \'areas como las de ciencias m\'edico
biol\'ogicas.

Se espera que el software interactivo corresponda a una herramienta
valiosa en el mejoramiento de la habilidad y capacidad de los
estudiantes de f\'isica para resolver ejercicios que son calificados
por una computadora y que por lo general cuentan con un tiempo
limitado para resolver los ex\'amenes. Tambi\'en se desea que el
alcance de la informaci\'on contenida en este software sea evidente,
ya que est\'a planeado para ser puesto en l\'inea, lo que permitir\'a
que en menor tiempo y en cualquier lugar sea distribuida y
consultada. Con esto se espera que los estudiantes puedan tener un
acceso remoto y un acercamiento f\'acil al conocimiento de la f\'isica
en varias de sus ramas de estudio.

\section*{Agradecimientos}\label{agradecimientos}

El autor agradece a la Facultad de Ciencias de la UNAM donde esta
propuesta fue concebida, y al Instituto Tecnol\'ogico de Kanazawa en
Jap\'on donde este trabajo fue comenzado y desarrollado y donde el
autor realiz\'o una estancia de investigaci\'on en c\'omputo para el
desarrollo de un sistema interactivo para el \'area de f\'isica
computacional.

\bibliographystyle{unsrtnat}
\bibliography{references}

\end{document}